\documentclass[a4paper,11pt]{article}
\usepackage{pos}

\usepackage{physics}
\newcommand{\NNN}{\nonumber\\ &}
\usepackage{graphicx}
\usepackage{amsmath}
\usepackage{amsfonts}
\usepackage{float}
\usepackage{hyperref}
\newcommand{\vect}[1]{\boldsymbol{#1}}

\title{Progress on computing the hadronic vacuum polarization contribution to the muon anomalous magnetic moment with staggered fermions}
 \ShortTitle{Progress on the HVP contribution to muon $g - 2$}

\author*[a]{Vaishakhi Moningi}
\author[b]{Christopher Aubin}
\author[a]{Thomas Blum}
\author[c]{Maarten Golterman}
\author[a]{Luchang Jin}
\author[d]{Santiago Peris}

\affiliation[a]{Dept. of Physics, Univ. of Connecticut, \\
Storrs, CT 06269, USA}

\affiliation[b]{Dept. of Physics \& Engineering Physics, Fordham Univ.,\\
Bronx, NY 10458, USA}

\affiliation[c]{Dept. of Physics and Astronomy, San Francisco State Univ., \\San Francisco, CA 94132, USA}

\affiliation[d]{Dept. of Physics and IFAE-BIST, Univ. Autònoma de Barcelona, \\
E-08193 Bellaterra, Barcelona, Spain}

\emailAdd{vaishakhi.moningi@uconn.edu}

\abstract{We give an update of our calculation of the light-quark, connected, hadronic vacuum polarization contribution to the muon anomalous magnetic moment, or muon $g-2$. The update includes preliminary results on a $2 + 1 + 1$ highly-improved staggered quark (HISQ) ensemble from the MILC collaboration with physical pion mass, $0.042$ fm lattice spacing, and volume $144^3 \times 288$. We discuss code and algorithm improvements for these calculations to compute the vector-vector correlation function more efficiently.}

\FullConference{The 41st International Symposium on Lattice Field Theory (LATTICE2024)\\
 28 July - 3 August 2024\\
Liverpool, UK\\}

\begin{document}
\maketitle

\section{Introduction}
The anomalous magnetic moment of the muon, $a_\mu = (g-2)/2$, can be measured and computed from theory very precisely. We can potentially uncover new particles and/or interactions by comparing the results from high-precision experiments and theory. The most precise experimental result right now is from Fermilab experiment E989 with a precision of $0.20$ ppm \cite{aguillard2023measurement}. Runs 4, 5, and 6 will further reduce the uncertainty by roughly a factor of 2. The result is anticipated for early 2025. Therefore, a commensurate theoretical effort is necessary to match the increasing experimental precision.

In this proceedings we provide an update on the leading hadronic vacuum polarization (HVP) contribution to muon $g - 2$ coming from the light quarks only using the HISQ action for 2+1+1 flavors of quarks. In particular we investigate two techniques to reduce statistical errors.

\section{Theoretical framework}
The total HVP contribution to $a_\mu$ comes from both connected and disconnected quark diagrams 
for each quark flavor. The largest contribution is from the connected light quarks, and we focus on those contributions in this work. Using lattice QCD and continuum, infinite-volume perturbative QED, one can calculate the HVP contribution to the muon anomalous magnetic moment~\cite{Lautrup:1971jf,deRafael:1993za,Blum:2002ii}.

\begin{eqnarray}
\label{eq:amu}
a_\mu^{\rm HVP} &=& 4\alpha^{2}\int_{0}^{\infty}d q^2 \, f(q^2)\,{\hat\Pi}(q^2),\\
f(q^2) &=& \frac{m_\mu^2 q^2 Z^3(1-q^2 Z)}{1+m_\mu^2 q^2 Z^2},~~~Z ~=~ -\frac{q^2-\sqrt{q^4+4 m_\mu^2 q^2}}{2 m_\mu^2 q^2}.
\end{eqnarray}
$m_\mu$ is the muon mass, and $\hat\Pi(q^{2})$ is the subtracted HVP, $\hat\Pi(q^{2})=\Pi(0)-\Pi(q^{2})$, computed directly on a Euclidean space-time lattice from the Fourier transform of the vector current two-point function,
\begin{eqnarray}
\label{eq:ft}
\Pi^{\mu\nu}(q) &=& \int d^4x\, e^{i q x}\langle j^\mu(x)j^\nu(0)\rangle = \Pi(q^2)(-q^\mu q^\nu+q^2\delta^{\mu\nu}) \label{eq:wi},\\
j^\mu(x)&=&\sum_{i} Q_i\bar\psi_i(x)\gamma^\mu\psi_i(x).
\end{eqnarray}
$j^\mu(x)$ is the electromagnetic current, and $Q_i$ is the quark electric charge in units of the electron charge $e$ and $i$ is the sum over flavours.
The form in Eq.~(\ref{eq:wi}) is found through the Lorentz invariance and the Ward identity.

To compute $a^{\rm HVP}_\mu$ we use the time-momentum representation~\cite{Bernecker:2011gh} which results from interchanging the order of the Fourier transform and momentum integrals in Eqs.~(\ref{eq:amu}) and (\ref{eq:ft}), respectively{:}
\begin{eqnarray}
\label{eq:hvp}
\Pi(0) -\Pi(q^2) &=& \sum_t \left(
\frac{\cos{qt} -1}{q^2} +\frac{1}{2} t^2
\right) C(t),~~~\label{eq:corr} C(t) ~ = ~\frac{1}{3}\sum_{\vec x,i}\langle j^i(\vec x, t)j^i(0)\rangle ,\\
\label{eq:kernel}
 w(t) &=& 4 \alpha^2 \int_{0}^{\infty} {d \omega^2} f(\omega^2)\left[\frac{\cos{\omega t} -1}{\omega^2} +\frac{t^2}{2}\right],
\end{eqnarray} 
 where $C(t)$ is the Euclidean time correlation function of two electromagnetic currents, averaged over spatial directions to project onto zero spatial momentum. Equation~(\ref{eq:amu}) becomes
 \begin{eqnarray}
 \label{eq:t-m amu}
a_{\mu}^{\rm HVP}(T) &=& \sum_{{t=-T/2}}^{T/2}  w(t) C(t)=2\sum_{t=0}^{T/2}  w(t) C(t).
\end{eqnarray}
$T$ is the temporal size of the lattice, and $a_{\mu}^{\rm HVP}$ is obtained in the limit $T\to \infty$. In Fig.~\ref{fig:amu integrand} we show a typical result for $w(t)C(t)$. 

The aim of lattice calculations is to efficiently obtain $C(t)$ with as much precision as possible. To do this we use the techniques of low-mode averaging (LMA)~\cite{Giusti:2004yp,DeGrand:2004qw} and all-mode averaging (AMA)~\cite{Blum:2013xva}. Both methods rely on the spectral decomposition of the quark propagator in terms of the eigenvectors of the lattice Dirac operator. The quark propagator from source point $y$ to sink point $x$, $S(x,y)$ mathematically is given by the inverse of the Dirac operator,
\begin{eqnarray}
S(x,y)= D^{-1}(x,y) =\sum_{\lambda\leq\lambda_N}\frac{\langle x|\lambda\rangle\langle\lambda|y\rangle}{\lambda}
+\sum_{\lambda>\lambda_N}\frac{\langle x|\lambda\rangle\langle\lambda|y\rangle}{\lambda}=
S_L+S_H,
\label{eq:spectral}
\end{eqnarray}
where the RHS shows the spectral decomposition divided into two pieces, one for the low modes up to {$\lambda_N$} and one for the rest, or high modes. Accordingly, we separate $C(t)$ into four parts,
\begin{equation}
    C(t)= \sum_{\vec x, \vec y} {\rm Tr}\gamma_i S(x,y)\gamma_i S(y,x)= C_{LL}+C_{LH}+C_{HL}+C_{HH},
\end{equation}
In practice, once the low modes are determined, $S_H$ is determined by computing
the inverse of the deflated Dirac operator, using the conjugate gradient algorithm. So, not only do the low modes significantly enhance our statistics through LMA, but they also dramatically accelerate the computation of the high mode part.

\section{Motivation}

\begin{figure}[h]
\begin{center}
\includegraphics[width=0.33\textwidth]{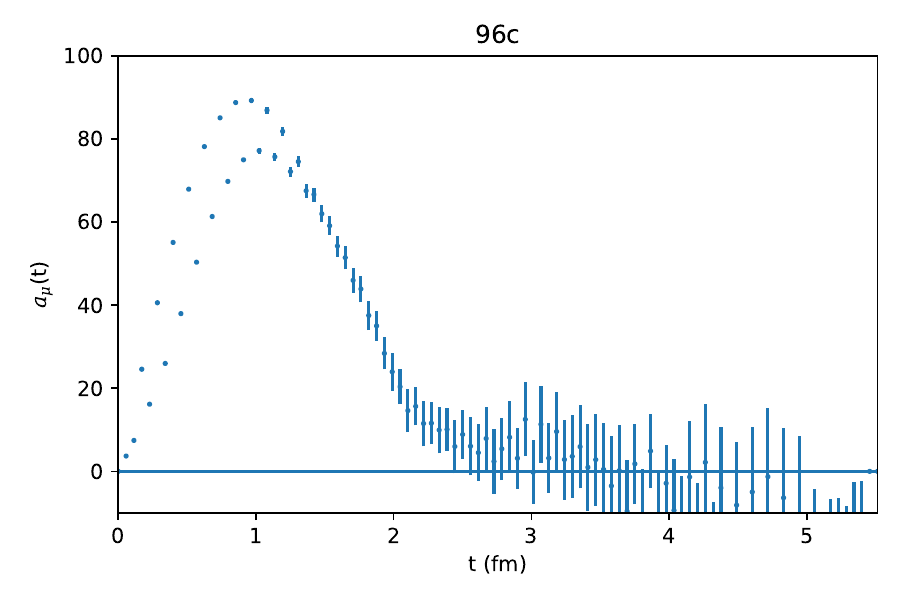}\hskip .0\textwidth
\includegraphics[width=0.33\textwidth]{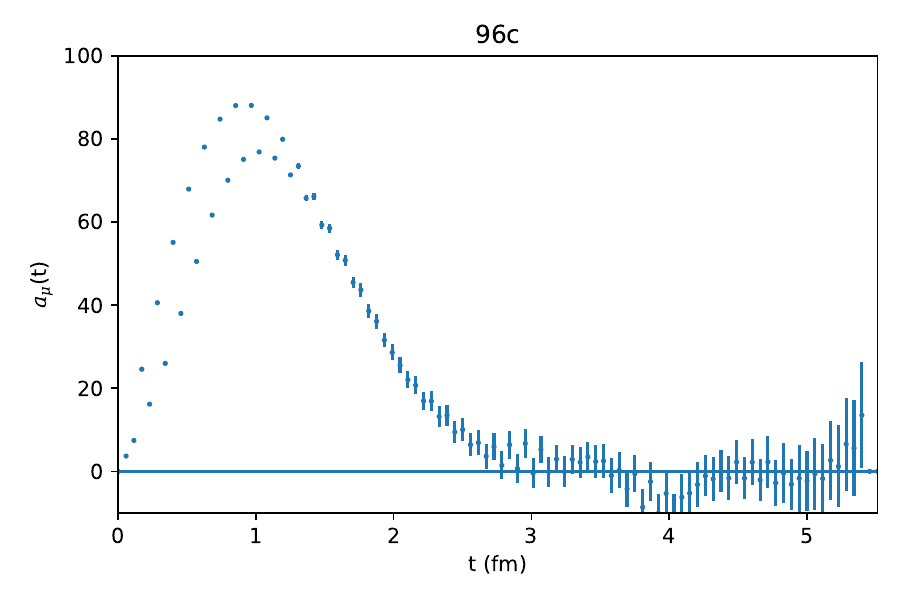}
\includegraphics[width=0.33\textwidth]{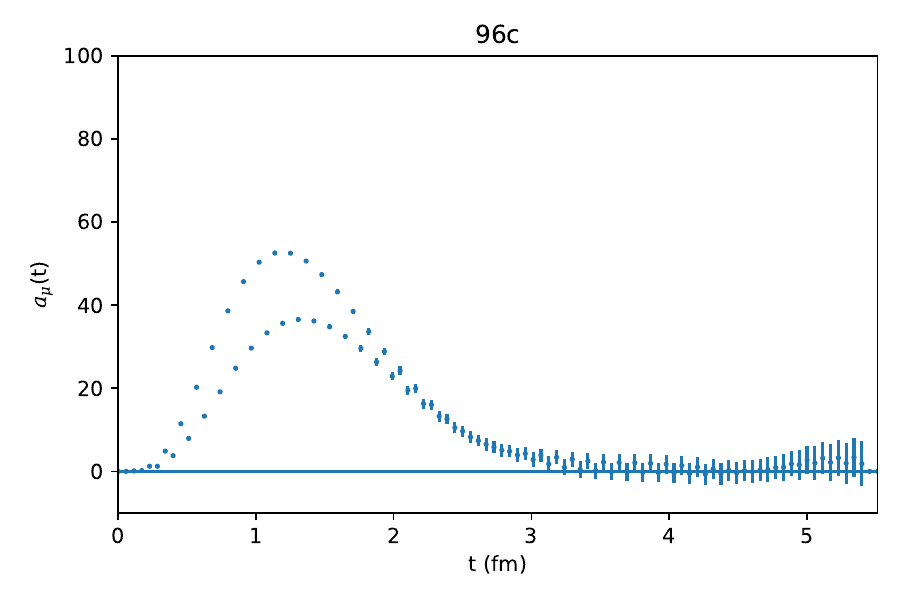}
\caption{The summand in Eq.~(\ref{eq:t-m amu}). No LMA (left), total (middle), LMA only (right).
Odd-parity, excited state oscillations intrinsic to staggered fermions are readily apparent.}
\label{fig:amu integrand}
\end{center}
\end{figure}

In our previous work \cite{Aubin:2019usy,Aubin:2022hgm}, the Euclidean correlation function $C(t)$ was just divided into two pieces: the pure low-mode contribution and the rest.
 We relied on LMA to handle the noisy long distance part of the
correlator and AMA for the rest. While this worked well for the result shown in Fig. \ref{fig:amu integrand},
we must do better to reach our goal, especially since the lattice of our target ensemble is even
larger, $144^3 \times 288$. First, while the LL part of the correlation function yields a full-volume
average over all source points, the HL and LH parts are only averaged over a tiny fraction
of them as part of the AMA procedure. Despite the exponentially small contribution at
long distance, the noise coming from the low modes in the HL (LH) part is still significant.
Comparing the middle and right panels in Fig. \ref{fig:amu integrand}, the LL part clearly has much smaller fluctuations than the total. 

To include a full volume average for the HL (LH) part, we adopt the method in~\cite{Giusti:2004yp} where the HL part is computed separately instead of together with the HH part as in our earlier AMA-style calculation~\cite{Aubin:2022hgm}\footnote{We thank Simon Kuberski and the Mainz group for discussions on this point.}. Note, if we did not do this separation, we would need more solves for the “rest” since then the high-high and high-low parts are computed all at once (an extra subtraction of the low-low
parts for these solves is also needed).
\section{Algorithm improvements}
On the lattice the local electromagnetic current is not conserved, so we use a point-split current that is exactly conserved,
\begin{align}
    J^\mu(x) &= -\frac{1}{2}\eta_\mu(x)(\Bar{\chi}(x+\hat{\mu})U^\dagger_\mu(x)\chi(x)+\Bar{\chi}(x)U_\mu(x)\chi(x+\hat{\mu}))
    \label{eq:current}
\end{align}
where $U_\mu$ are the gauge links for gauge invariance, $\chi(x)$ are the single spin component staggered fermion fields, and $\eta(x)$ arise from the spin diagonalization of the fermion action. 
The two-point function then becomes
\begin{align}
    \langle J^\mu(x)(J^\nu(y))^\dagger\rangle &= \frac{1}{4}(S(y+\hat{\nu},x+\hat{\mu})\eta_\mu(x)U^\dagger_\mu(x)S(x,y)\eta_\nu(y)U_\nu(y)\NNN+ S(y+\hat{\nu},x)\eta_\mu(x)U_\mu(x)S(x+\hat{\mu},y)\eta_\nu(y)U_\nu(y)\NNN+S(y,x+\hat{\mu})\eta_\mu(x)U^\dagger_\mu(x)S(x,y+\hat{\nu})\eta_\nu(y)U^\dagger_\nu(y)\NNN+S(y,x)\eta_\mu(x)U_\mu(x)S(x+\hat{\mu},y+\hat{\nu})\eta_\nu(y)U^\dagger_\nu(y)).
    \label{eq:2pt_pt}
\end{align}
 For the low-mode contributions it is useful to define a quantity called a meson field, 
\begin{align}
        \left(\Lambda_\mu(t)\right)_{n, m}=\sum_{\vec x}\expval{n|x} U_\mu(x)\expval{ x+\mu | m},
    \label{eq:mf}
\end{align}
where $n,m$ label eigenmodes of the Dirac operator, and the phases have been absorbed into the link variables, $U_\mu(x)\eta_\mu(x)\to U_\mu(x)$.
\subsection{High-low (HL) contribution} 

Substituting the low-mode part of the spectral decomposition of one of the two quark propagators, the HL part of the two-point current-current correlation function then becomes
\begin{align}
   C_{HL} &=\frac{1}{4} \sum_{ n}\sum_{\vect{y}}\left[\braket{n}{y}U_\nu(y)S(y+\hat{\nu},x+\hat{\mu})U^\dagger_\mu(x)\frac{\braket{x}{n}}{\lambda_n} + \braket{n}{y}U_\nu(y)S(y+\hat{\nu},x)U_\mu(x)\frac{\braket{x+\hat{\mu}}{n}}{\lambda_n}\right. \NNN
 + \left.\braket{n}{y+\hat{\nu}}U^\dagger_\nu(y)S(y,x+\hat{\mu})U^\dagger_\mu(x)\frac{\braket{x}{n}}{\lambda_n} + \braket{n}{y+\hat{\nu}}U^\dagger_\nu(y)S(y,x)U_\mu(x)\frac{\braket{x+\hat{\mu}}{n}}{\lambda_n}\right]
   \label{eq:HL}
\end{align}
To effect the full volume average, each low mode is used as a source for the high mode part of the propagator in Eq. (\ref{eq:HL}). The price to pay is one quark propagator inversion for each eigenvector on each time slice, or a total of $N_T\times 2N_{\rm{low}}$ sources. Note $N_{\rm{low}}$ is the number of preconditioned Dirac operator low-modes, and there are two low-modes of the full Dirac operator for each. To dramatically reduce the cost, we form a linear combination of the low-mode sources on a given time-slice using a unique random number for each and contract at the sink with the same random number which kills the unwanted cross-terms on average. While this adds random noise to the calculation, we find the average over low-modes still wins to reduce the final statistical noise. Furthermore we can systematically reduce the noise by doing more ``hits'' with additional random-sources.

\begin{table}[htbp]
\centering
\resizebox{0.8\textwidth}{!}{
\begin{tabular}{|l|l|l|l|l|l|l|}
\hline
$m_\pi$ (MeV) & $a$ (fm)  & size              & $L$ (fm)     & $m_\pi L$ & $N_{\text{low}}$   & \begin{tabular}[c]{@{}l@{}}\# configs\\ (LL-HL-HH)\end{tabular} \\ \hline
130          & 0.08787 & $64^3\times 96$   & 5.62  & 3.66      & 4000 & 31-31-31                                                     \\ \hline
134          & 0.042   & $144^3\times 288$ & 6.048 & 3.95      & 4000 & 6-13-17                                                      \\ \hline
\end{tabular}%
}
\caption{ Lattice simulation details. “$N_{\text{low}}$” is the number of low-modes of the preconditioned Dirac operator. The number of configurations used for measurements in this study are given in the last column. The HISQ 2+1+1 flavor ensembles used here were produced by the MILC collaboration.}
\label{tab:config}
\end{table}

We use configurations generated by the MILC Collaboration \cite{Bazavov:2014wgs,Bazavov:2017lyh} with 2+1+1 flavors of HISQ
quarks. The new calculations are done on the two ensembles shown in Table \ref{tab:config}. The first was used in our previous calculations~\cite{Aubin:2019usy,Aubin:2022hgm} and the second is a new one with lattice spacing 0.042 fm, which is the finest to date. Both have roughly the physical pion mass, and we have computed 8000 low-modes of the full Dirac operator on each.

\subsection{Low-low (LL) contribution}
The LL correlation function is easily constructed using the meson fields in Eq.~(\ref{eq:mf}),
\begin{align}
    C_{LL}= \frac{1}{4} \sum_{m, n} 
    \sum_{\vec{x}} \frac{1}{\lambda_m\lambda_n}&\left[ \Lambda^\dagger_\mu(x)_{mn}\Lambda^\dagger_\nu(y)_{nm} + \Lambda^\dagger_\mu(x)_{mn}\Lambda_\nu(y)_{nm} \right. \NNN+ \left. \Lambda_\mu(x)_{mn}\Lambda^\dagger_\nu(y)_{nm} + \Lambda_\mu(x)_{mn}\Lambda_\nu(y)_{nm} \right],
\end{align}
where $\lambda_n$ is shorthand for $i\lambda_n+m$.
The cost of the meson field scales linearly in the size of the lattice and quadratically with the number of eigenvectors which is prohibitive when both are large as in the case of the $144^3\times288$ lattice.

To significantly speedup the calculation, we “sparsen” the eigenvectors by omitting some number of sites in a regular pattern, $i.e.$, omit $s$ consecutive points, evenly spaced, in each direction. 
This is increasingly effective as $a\to 0$ since nearby points will be more and more correlated, and including them in the average does not meaningfully improve the statistical error. Sparsening also drastically reduces the memory footprint which is important if later we decide to increase the number of eigenvectors. 

To ensure we preserve the spin-taste structure of the staggered fermion currents, the sparsening is done by choosing the location for a {\it hypercube} randomly on a given time-slice, and then we omit every $s$ number of hypercubes in each spatial direction. In other words we always keep all points in a kept hypercube. By randomly choosing the initial hypercube on a time-slice we are guaranteed to project onto zero momentum in the average. Sparsening by a factor $(s,t)$ reduces the size of eigenvectors required to compute our meson fields from $N_{\rm{S}}^3 \times N_{\rm{T}}$ to $(N_{\rm{S}}/s)^3 \times (N_{\rm{T}}/t)$.
\vspace{-0.cm}
\subsection{High-high (HH) contribution}
\vspace{-0pt}
The HH part of the correlation function dominates the small Euclidean time regime so has exponentially smaller errors compared to the long-distance part. We compute the HH part on a regular, sparse, grid of point sources, as before~\cite{Aubin:2019usy,Aubin:2022hgm}, except now we subtract off the low-mode part. 
Finally, all parts of the correlation function can be performed inside an AMA framework where most of the components are computed approximately, and therefore relatively cheaply, and corrected with exact (to numerical precision) solves infrequently~\cite{Blum:2012uh}.
\section{Results and Conclusions}
\begin{figure}[H]
        \centering
        \includegraphics[width=0.9\textwidth]{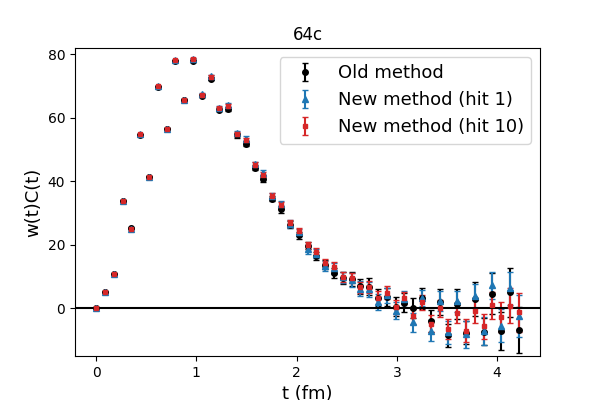}
        \caption{Comparision of the summand in Eq.~\ref{eq:t-m amu} between our old method and new method for the $64^3$ ensemble.}
        \label{fig:64c comparision}
    \end{figure}

\begin{figure}[H]
        \includegraphics[width=7.8cm]{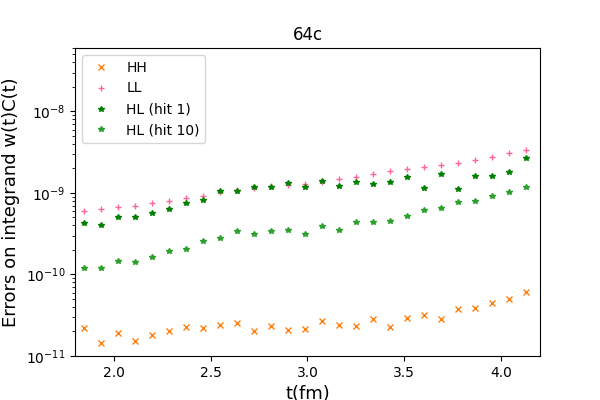}   
         \includegraphics[width=7.8cm]{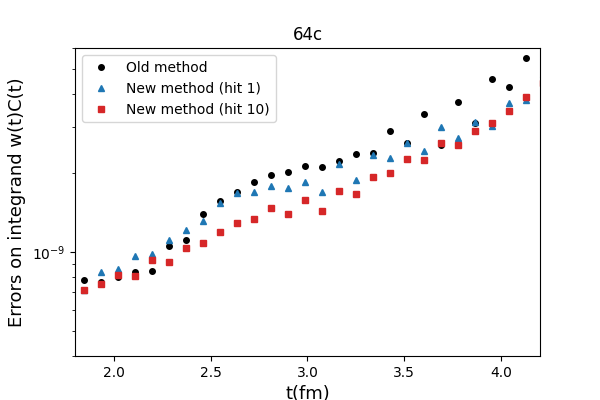} 
         \caption{Statistical errors on the summand in Eq.~(\ref{eq:t-m amu}). Comparison between number of hits (left) and comparison between old and new method (right). }
        \label{fig:64c err Total}
\end{figure}

For the $64^3$ ensemble, we compare our older results for the summand to the newer results in
Fig. \ref{fig:64c comparision}. 
The new method reduces the number of solves from $N_{\rm{T}}\times 2N_{\rm low}$ to $N_{\rm{T}}\times N_{\rm hits}$. One can see the reduction in errors in the long-distance region with the new data, especially for 10 hits. The error contributions from the HL part of the correlation function appear to be significantly suppressed from the LL part (see Fig.~\ref{fig:64c err Total}). Our current method, which involves using a “1 hit” approach, has demonstrated an improvement of $6.78$\% for the total error compared to the old method. This improvement becomes more pronounced when employing a “10 hits” approach, yielding a $23.7$\% improvement within the long-distance window of $2.3-3.3$ fm. 

\begin{figure}[H]
        \centering
        \includegraphics[width=8cm]{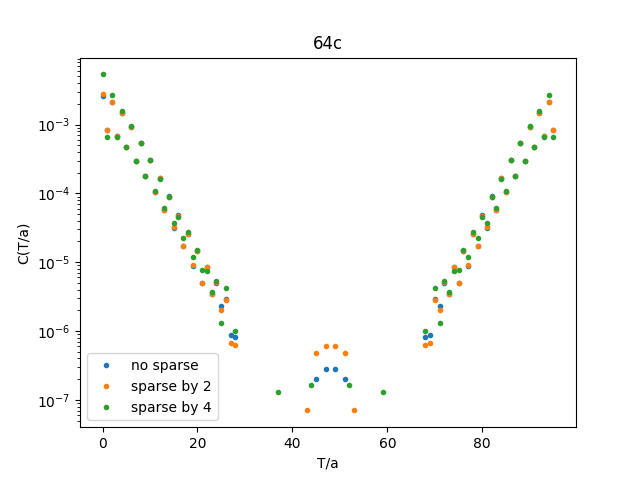}
        \caption{LL contribution from contracting the meson fields with, and without, sparsening the low-modes.}
        \label{fig:LL-sparse}
    \end{figure}

We emphasize that the LL part remains the dominant source of uncertainty in the correlation function calculation. As mentioned earlier, to minimize the cost of computing the LL part, we implement sparsening to generate the meson fields. For demonstration purposes, we have used 800 low modes on the $64^3$ lattice for one configuration (see Fig. \ref{fig:LL-sparse}). By applying sparsening with a factor of 2(or 4) in each spatial dimension, the number of sites at both the source and sink is reduced by a factor of 8(or 64). This reduction lead to a substantial increase in computational speed for the meson fields. 
Although these techniques show significant potential, further studies and research are required to fully assess the trade-offs involved and the broader applicability of this approach.  
\subsection{Preliminary Results on the $144^3\times 288$ lattice}
\begin{figure}[H]
    \centering
      \includegraphics[width=7.5cm]{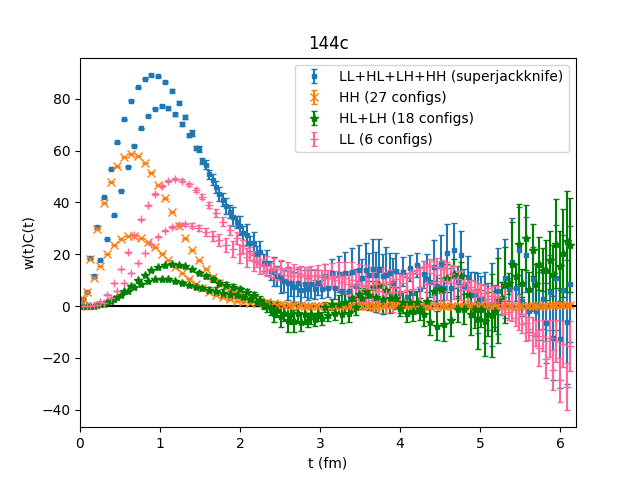}
      \includegraphics[width=7.5cm]{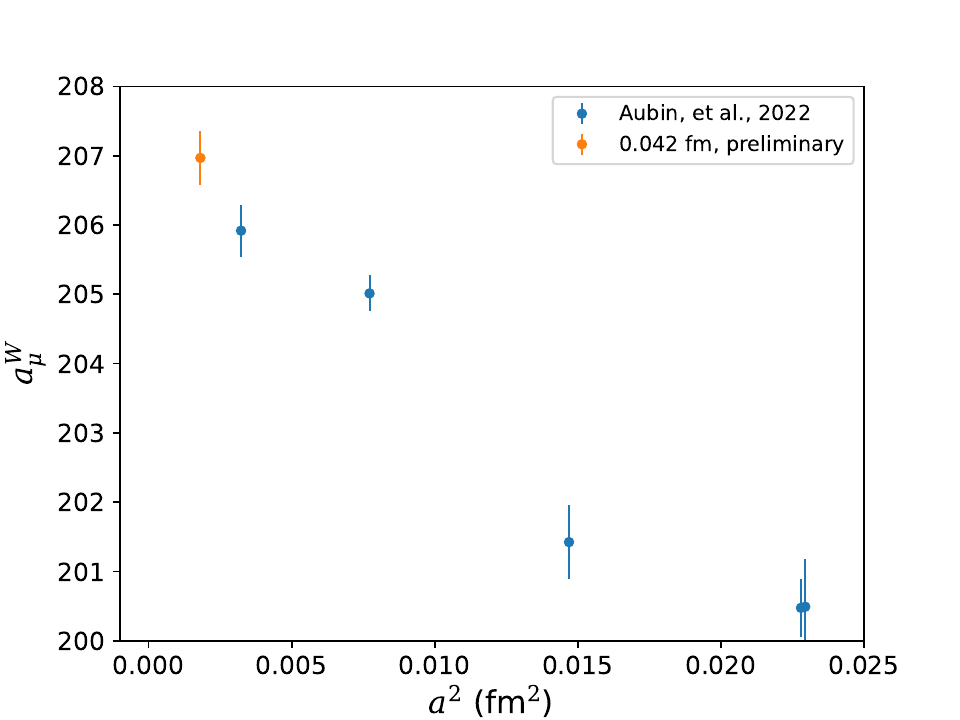} 
       \caption{Summand in Eq. (\ref{eq:t-m amu}) for the $144^3$ ensemble (left) and uncorrected results for the intermediate window value. The new point at $a=0.042$ fm (see Table \ref{tab:144c intermediate win}) is compared with our earlier results~\cite{Aubin:2022hgm} (right). Errors shown are statistical only.}
        \label{fig:144c}
\end{figure}

Simulations on finer lattices are needed to determine whether the R-ratio and lattice values are consistent. As a first step we have computed preliminary values of $a_\mu^{\rm HVP}$ for two windows~\cite{Blum:2018mom}, which are given in Tab.~\ref{tab:144c intermediate win}. We have implemented the new HL method, but not yet the sparsening of the LL part. For the HL part, we use one hit on each of 288 time-slices rather than more hits on fewer time slices. For the HH part, we use $4^3$ point sources on eight time slices, or 512 sources evenly spread over the lattice volume. The summand in Eq.~(\ref{eq:t-m amu}) is shown in Fig.~\ref{fig:144c} (left panel).
%
\begin{table}[H]
\centering
\resizebox{0.3\textwidth}{!}{%
\begin{tabular}{|l|l|}
\hline
$a_\mu^{\rm HVP, win} \times 10^{10}$ & window ($t_0,t_1,\Delta$) (fm) \\ \hline
206.9(45)          & (0.4, 1.0, 0.15)               \\ \hline
94.6(4.16)         & (1.5,1.9,0.15)                 \\ \hline
\end{tabular}%
}
\caption{$a_\mu^{\rm HVP, win} \times 10^{10}$ for two different windows computed on the $144^3$ ensemble. Errors are statistical.}
\label{tab:144c intermediate win}
\end{table}
The (uncorrected) results for the windows in Tab.~\ref{tab:144c intermediate win} appear promising, even with few measurements. Calculations on more configurations are proceeding which will provide a critical test of the new methods for the long-distance window.
Although the method requires the added effort of computing HL separately, this trade-off is justified by the reduction of errors following the full volume average of the low-modes. Moreover, the use of fewer sources for the HH part, now free from the additional noise contributed by the HL part, enhances computational efficiency. Likewise, sparsening the LL part leads to a dramatic speedup. These improvements are expected to accelerate computations, particularly on finer lattices as our work progresses with the $144^3$ ensemble.


\acknowledgments
TB and VM were partially supported by the US DOE Office of Science under grant DE-SC0010339. LJ was supported by the DOE Office of
Science Early Career Award DE-SC002114. MG is supported by the U.S.\ Department of Energy,
Office of Science, Office of High Energy Physics, under Award No.
DE-SC0013682. SP is supported by the Spanish Ministerio de Ciencia e Innovacion,
grants PID2020-112965GB-I00 and PID2023-146142NB-I00,
 and by the Departament de Recerca i Universities from Generalitat de Catalunya
 to the Grup de Recerca 00649 (Codi: 2021 SGR 00649).
IFAE is partially funded by the CERCA program of the Generalitat de Catalunya.

The authors acknowledge the \href{ http://www.tacc.utexas.edu}{Texas Advanced Computing Center} (TACC) at The University of Texas at Austin for providing computational resources that have contributed to the research results reported within this proceedings.
\bibliographystyle{JHEP}
\bibliography{ref.bib}

\end{document}